# DPTraj-PM: Differentially Private Trajectory Synthesis Using Prefix Tree and Markov Process

Nana Wang

Jiangsu Normal University, School of Computer Science and Technology, wangnana_5@aliyun.com

Mohan Kankanhalli

National University of Singapore, School of Computing, mohan@comp.nus.edu.sg

The increasing use of GPS-enabled devices has generated a large amount of trajectory data. These data offer us vital insights to understand the movements of individuals and populations, benefiting a broad range of applications from transportation planning to epidemic modeling. However, improper release of trajectory data is increasing concerns on individual privacy. Previous attempts either lack strong privacy guarantees, or fail to preserve sufficient basic characteristics of the original data. In this paper, we propose DPTraj-PM, a method to synthesize trajectory dataset under the differential privacy (DP) framework while ensures high data utility. Based on the assumption that an individual's trajectory could be mainly determined by the initial trajectory segment (which depicts the starting point and the initial direction) and the next location point, DPTraj-PM discretizes the raw trajectories into neighboring cells, and models them by combining a prefix tree structure and an $m$-order Markov process. After adding noise to the model under differential privacy, DPTraj-PM generates a synthetic dataset from the noisy model to enable a wider spectrum of data mining and modeling tasks. The output traces crafted by DPTraj-PM not only preserves the patterns and variability in individuals' mobility behaviors, but also protects individual privacy. Experiments on two real-world datasets demonstrate that DPTraj-PM substantially outperforms the state-of-the-art techniques in terms of data utility. Our code is available at https://github.com/wnn5/DP-PrefixTreeMarkov.

**ACM Reference Format:**
Nana Wang and Mohan Kankanhalli. 2024. DPTraj-PM: Differentially Private Trajectory Synthesis Using Prefix Tree and Markov Process.. *1, 1,* Article 1, 23 pages.

## 1 INTRODUCTION

Nowadays, the popularity of GPS (Global Positioning System)-enabled devices and the rise of ubiquitous wireless connectivity have generated a wealth of personal trajectory data (Figure 1). These data, representing the mobility behavior of individuals, could be used for many data mining and modeling applications [1-2], such as traffic flow analysis, urban planning, marketing analysis, and epidemic modeling. On the other hand, the trajectory data are highly sensitive [3-5]. Home and work locations, health status, religious beliefs, and personal relationships can be easy to infer from the individuals' whereabouts [2]. Due to the concern of privacy leakage, companies and researchers are reluctant to release datasets. This hinders many data-driven studies from analyzing such data to best serve the general public.

Many research efforts have been made to prevent privacy leakage [6-14]. For example, [8] prevent users' trajectories from being identified by making the trajectories $k$-anonymized; [9-10] introduce distortions to the trajectories by disturbing the connectivity of trajectory segments or the temporal information; [11] generates dummies to lower the identification disclosure rate; [12] synthesizes new trajectories via deep learning models; [13-14] protect sensitive place visits based on semantic cloaking. These methods can protect privacy under specific attack models. But once the adversary can gather more background knowledge, sensitive information of individuals may still be disclosed [1-2, 15].

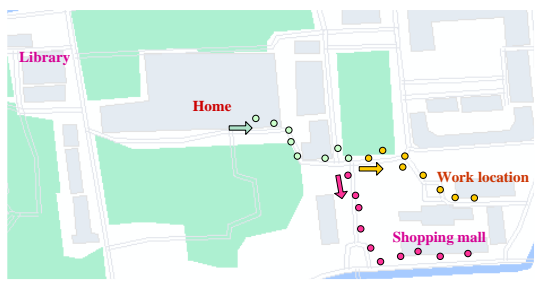

Figure 1: An example of two GPS raw trajectories.

Recently, differential privacy (DP) [16] has become the leading criterion that can provide provable and strong guarantees of privacy. The main idea of differential privacy is to add noise with a privacy budget $\varepsilon$ to a dataset such that an adversary cannot decide whether a particular record is included in the dataset or not [17]. In general, the solutions implementing such privacy principle construct some mobility models (e.g., trees [17-19], or probability distributions [20-23]) which summarize the individuals' complete movement behaviors from the original data, apply noise so as to make the model differentially private, and then generate synthetic trajectories from the noisy model [2]. However, due to the inherent sequentiality and high-dimensionality of trajectory data, it is still a challenge to both preserve the characteristics of the real data and fully meet the differential privacy criterion. To illustrate our claim, we compare our method DPTraj-PM with four most relevant works: DPT [19], AdaTrace [20], DP-MODR [21] and LDPTrace [22] in Figure 2. The probability distribution of visited top-$n$ regions in the generated trajectories for each technique has been compared to the true dataset (Taxi-1 [24]). We select $n = 20$ out of a total of 400 regions. The red curve shows the visit probability distribution for the true dataset. The blue dash curve represents the trend line of the visit probability distribution for the synthetic dataset. Clearly, our approach has the closest fit to the true distribution, which means the top-$n$ regions are very well preserved in the synthetic dataset. In Section 5, its superiority is demonstrated also quantitatively in terms of other 6 metrics.

In this paper, we present DPTraj-PM, a trajectory synthesizer with both high data utility and differential privacy. Our key idea is quite intuitive. As shown in Figure 1, if a person starts from his home and goes along the way indicated by the light green arrow, he has a much higher probability of going to his work location or the shopping mall than the library. When the initial trajectory segment in light green is finished, his next location could be predicted by previous locations. Therefore, we build a mobility model from the trajectories under the assumption that an individual's trajectory could be mainly determined by the initial trajectory segment (which depicts the starting point and the initial direction) and the next location point. To achieve this goal, DPTraj-PM discretizes the raw trajectories into neighboring cells, and organizes them by combining a prefix tree structure and an $m$-order Markov process. By carefully designing the added noise under differential privacy for this model, DPTraj-PM can generate a synthetic dataset with high data utility. We theoretically prove that DPTraj-PM satisfies differential privacy. The experiments on two real-world datasets show that DPTraj-PM significantly outperforms existing algorithms in general in terms of data utility.

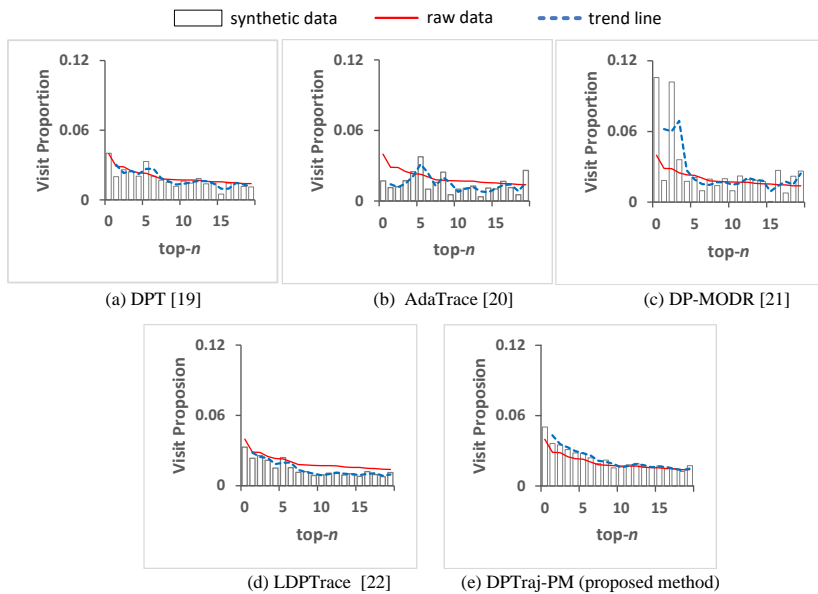

Figure 2: Probability distribution of top-*n* visited regions for real and synthetic trajectories generated by each method over Taxi-1 ($\varepsilon = 1$): (a) DPT [19], (b) AdaTrace [20], (c) DP-MODR [21], (d) LDPTrace [22], and (e) DPTraj-PM (proposed method).

We need to point out that although some existing methods have adopted the prefix tree [17-19] or 1-order Markov process [20-23] for differentially private trajectory synthesis, DPTraj-PM is the first that models the trajectories using a prefix tree followed by an *m*-order Markov model (Figure 3), and adds carefully designed noise to both of them under differential privacy for privacy protection and better data utility. The prefix tree helps DPTraj-PM to preserve the initial directions and the starting points, while the *m*-order Markov process for the next location point prediction allows DPTraj-PM to save some privacy budget that is required by a high prefix tree, and to generate longer trajectories. Furthermore, given enough trajectory data and the same privacy budget allocated to a Markov process, predicting the next location point based on an *m*-order Markov process which relies on *m* previous location points could be more accurate than a 1-order Markov process.

The rest of the paper is organized as follows: Section 2 gives an overview of related work. Section 3 presents the preliminaries for understanding the proposed model. We detail DPTraj-PM in Section 4. Section 5 provides our experimental results of the algorithm. We conclude this paper in Section 6.

## 2 RELATED WORK

In this section, we review the state-of-the-art algorithms for differentially private trajectory publishing, i.e., the methods for the publication (or release) of trajectory databases that preserve the privacy of individuals based on differential privacy. Depending on the mobility model used for trajectory organization, we classify them into three broad categories and discuss relevant works under each category.

**Modeling trajectories as trees.** Most works of this category deal with the trajectories that are represented by places [17-18, 25-28]. The first work to adopt differential privacy for such trajectories is that by Chen et al. [18]. They organize the trajectories as a prefix tree, i.e., a hierarchical framework which is built by grouping trajectories with the same prefix.

Each node of the prefix tree stores the count of a place sequence. Because of the unicity of trajectories, the counts for leaf nodes may be very sparse, and thus the data utility may be not desirable. In a subsequent work [17], sub-trajectories which are represented by variable-length *n*-grams are used to build an exploration tree. The leaf counts become higher, but the starting and ending regions of the trajectories may not be well preserved. For better data utility, Wang and Sinnott [26] build a noise-enhanced prefix tree by exploiting an adaptive pruning technique and a geometric privacy budget distribution strategy. To publish sanitized spatio-temporal trajectory data, Al-Hussaeni et al. [27] divide each level of a prefix tree into two sublevels: location and timestamp, according to a location taxonomy tree and a timestamp taxonomy tree. Li et al. [28] use an incremental privacy budget allocation technique and a spatial-temporal dimensionality reduction approach for a prefix tree to improve the data utility. Since each level consumes some privacy budget, the height of the prefix tree is limited, and thus it may still not scale well for long trajectories.

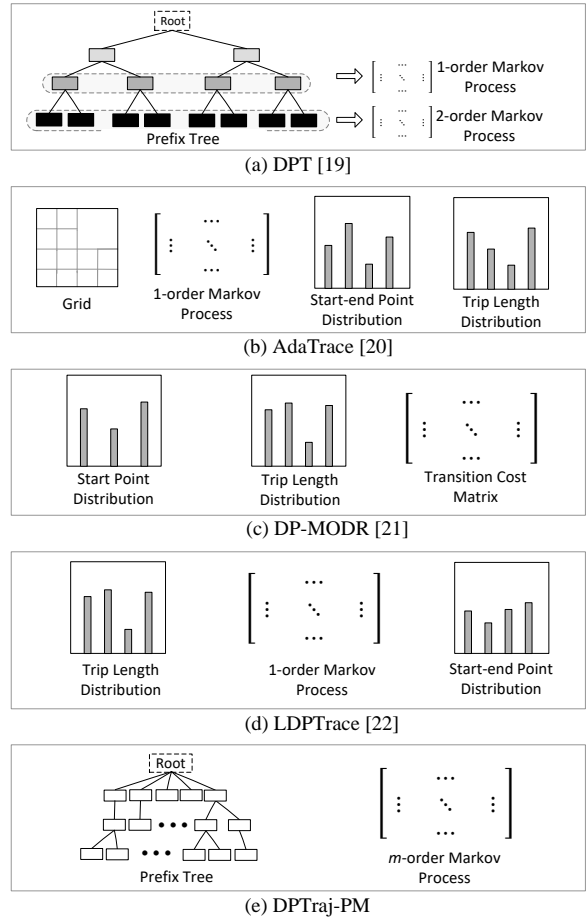

Figure 3: The main components of the models used by different methods: (a) DPT [19], (b) AdaTrace [20], (c) DP-MODR [21], (d) LDPTrace [22], and (e) DPTraj-PM.

**Modeling trajectories as clusters.** Algorithms of this category aim at dividing locations into clusters at each time point under differential privacy, and then drawing samples from the noisy clusters for synthetic trajectory release [29-31]. For example, Hua et al. [29] probabilistically merge locations at each time point via an exponential mechanism, and then adopt a noisy counting technique adapted from the Laplace mechanism to publish synthetic trajectories. Since the centroid of each merged location set is released without any modification, reconstructing some trajectories could be possible [1]. Zhao et al. [30] calculate the noisy clusters by adding Laplace noise into the location cluster centers at each time stamp. Liu et al. [31] adopt the Staircase mechanism [32] for the generation of the noisy cluster centers. Because each time stamp requires some privacy budget, these methods [29-31] may be not suitable for releasing long trajectories.

**Modeling trajectories as probability distributions.** This kind of methods [20-23, 33-37] rely on adding noise to a set of probability distributions that depict some key statistical features of the movements in the original trajectories, and generating synthetic trajectories from the noisy distributions. For instance, Mir et al. [33] use this methodology for the release of the Call Detail Records (CDRs) from a cellular telephone network. They add Laplace noise to five probability distributions including the spatial distribution of home and work locations, commute distance distribution, calls per day per user distribution, call times per user class distribution, and hourly calls per location distribution, for the release of synthetic trajectories. Roy et al. [34] group the attributes of a bicycle trajectory dataset, and add Laplace noise to the distributions for each group for synthetic trajectory generation. Moreover, Jin et al. [35] protects the trajectory privacy by perturbing the local/global frequency distributions of significantly important trajectory points. Only protecting some important points could limit its application when every location visit is regarded as highly sensitive. Rather than using the empirical distributions like [33-35], Ho et al. [37] use a three-player GAN (Generative Adversarial Network) [38] to learn the underlying distribution of consecutive data that meet the constraints of differential privacy, and generate a new dataset from the trained GAN. It offers high data utility, but may have a potential risk of compromising privacy, because it enforces differential privacy guarantees by training a resource-intensive learning algorithm (i.e., the three-player GAN).

**Most relevant work.** The methods in [19-22] are most relevant to our work. He et al. [19] present DPT (Differentially Private Trajectories), a system that exploits a hierarchical reference system to discretize the spatial domain at multiple resolutions, and builds a set of prefix trees, each referring to a different spatial resolution, for trajectory synthesis. Different from the prefix tree in [18], DPT builds each prefix tree by grouping the trajectory segments' variable-length $n$-grams with the same prefix into the same branch for each reference system. The $n$-grams of the starting segments are treated the same as any other ordinary $n$-gram, and are allocated with the same privacy budget as the ordinary $n$-grams at the same level. The direction indication ability of the starting segments may not be well preserved. Besides, due to the consistency enforcement operation, the noise added to lower levels could be propagated to higher levels (Figure 3(a)), and hence affects the corresponding Markov processes, which might lead to more data utility loss. Gursoy et al. [20] design AdaTrace, a method that uses four spatial and statistical features, which include a density-aware grid, a 1-order Markov mobility model, a start-end point distribution and a trip length distribution, to generate synthetic trajectories. Given the starting cell and the destination cell of a trip, a combination of 1-step transition probabilities is used to calculate the whole trip in the trajectory synthesis step. In [21], Deldar and Abadi introduce DP-MODR, an approach that adopts three spatial and statistical features including start point distribution, a trip length distribution, a transition cost matrix, for trajectory release. Du et al. [22] present LDPTrace, a locally differentially private trajectory synthesis method, which takes three crucial patterns into account: a trip length distribution, a 1-order Markov mobility model, and a start-end point distribution.

Our method DPTraj-PM organizes the raw trajectories using a height-($m$ + 2) prefix tree [18] and an $m$-order Markov process (Figure 3(e)). The prefix tree is used to organize the initial trajectory segments, while the $m$-order Markov process is employed to pick the next location point. The $m$-order Markov process only needs to accommodate the noise added to itself. Besides, DPTraj-PM utilizes a decremental privacy budget allocation technique in the noise prefix tree construction for better initial trajectory segment imitation. In Section 5, we experimentally compare DPTraj-PM with [19-22], and demonstrate that DPTraj-PM provides better data utility.

## 3 PRELIMINARIES

### 3.1 Notation

Let $D = \{T^i | i = 0,1,…, |D|−1\}$ denote a raw trajectory dataset that has $|D|$ trajectories, and $T^i$ be the $i$-th trajectory of $D$.

**Definition 1** (Raw trajectory [27]). A raw trajectory $T^i = \{(loc_j^i, t_j^i) | j = 0,1, …, |T^i| − 1\}$ is a sequence of $|T^i|$ tuples, with each tuple consisting of a location point $loc_j^i$ and a time stamp $t_j^i$ ($t_j^i \leq t_{j+1}^i$). Its $j$-th location point $loc_j^i$ is further denoted by ($loc_j^i.x, loc_j^i.y$), where ($loc_j^i.x, loc_j^i.y$) refers to the longitude and latitude of $loc_j^i$.

In this work, we focus on the spatial features of the raw trajectories. Hence $T^i$ is given as $T^i = \{loc_j^i | j = 0,1, …, |T^i| − 1\}$ for the rest of the paper. The latitude-longitude coordinates of the locations are usually recorded in some continuous domain $Z$. To simulate a move from one location to another, we have numerous possibilities to consider during synthetic trajectory generation. To limit the size of the model which will be built later, one common way to analyze trajectories is via discretization of the spatial domain using a reference system [19].

**Definition 2** (Reference system [19]). A reference system (*RS*) is a set of anchor points $AP \subset Z$, associated with a mapping function $f: Z \rightarrow AP$.

### 3.2 Prefix Tree

A prefix tree [18] is a type of a tree structure that can represent a sequential dataset in a compact way. In a prefix tree, sequences with the same prefix are grouped into the same branch. Let $S = p_0 p_1 \cdots p_{|S|−1}$ be a sequence of $|S|$ symbols, and $S' = p_0' p_1' \cdots p_{|S'|−1}'$ be a sequence of $|S'|$ symbols. $S'$ is a prefix of $S$, denoted by $S' \preccurlyeq S$, if and only if $|S'| \leq |S|$ and $\forall 0 \leq i \leq |S'| − 1, p_i' = p_i$. For example, $p_1 p_2 p_3$ is a prefix of $p_1 p_2 p_3 p_4$, but $p_2 p_3$ is not.

**Definition 3** (Prefix tree [18]). A prefix tree *PT* of a sequential dataset $\widetilde{D}$ is denoted by a triplet $PT = (V, E, Root)$, where $V$ is the set of nodes, each corresponding to a unique prefix in $\widetilde{D}$; $E$ is the edge set, representing transitions between nodes; and *Root* denotes the virtual root of *PT*. Let *prefix*($v$, *PT*) be the unique prefix represented by the node $v$, which starts from *Root* to $v$.

Each node $v \in V$ stores a pair $(tr(\widetilde{D}, v), c(v))$, where $tr(\widetilde{D}, v)$ denotes the set of sequences in $\widetilde{D}$ having *prefix*($v$, *PT*), and $c(v)$ is a noisy version of $|tr(\widetilde{D}, v)|$ (i.e., the number of sequences in $tr(\widetilde{D}, v)$, $|tr(\widetilde{D}, v)| = 0, 1, 2, …, |\widetilde{D}|$). For example, $c(v)$ can be obtained by adding Laplace noise to $|tr(\widetilde{D}, v)|$.

The level of the node $v$ is the length of the path from *Root* to $v$ (i.e., *prefix*($v$, *PT*)). *Root* is at level zero. The height of the prefix tree *PT* is one more than the level of the deepest node in the tree. Figure 4 shows an example of a sequential dataset and its corresponding prefix tree, where each node $v$ is labeled with its symbol and $|tr(\widetilde{D}, v)|$, and the end of each sequence is indicated with a stopping symbol #.

### 3.3 *m*-order Markov Process

An $m$-order Markov process is a statistical model that can be used to model correlations between contiguous symbols in a regular sequence [19, 39]. It is based on the *Markov independence assumption* which states that the presence of a particular symbol in a sequence depends only on the previous $m$ symbols [19, 39].

Formally, given a sequence $S = p_0 p_1 \cdots p_{|S|−1}$, for every $m \leq j < |S| − 1$, using the *Markov assumption* we have

$$Pr[p_{j+1} = p \mid p_0 \cdots p_j] = Pr[p_{j+1} = p \mid p_{j−m+1} p_{j−m+2} \cdots p_j]. \tag{1}$$

The probability $Pr[p_{j+1} = p \,|\, p_{j-m+1}p_{j-m+2}\cdots p_j]$ is referred to as a transition probability of the *m*-order Markov process.

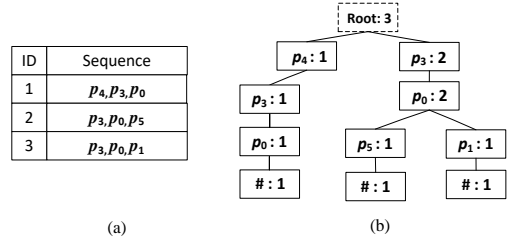

Figure 4: An example of a sequential dataset and its prefix tree: (a) a sequential dataset, and (b) the prefix tree corresponding to Figure 4(a).

Any sequence $x$ of $m$ symbols is called an *m*-gram. Let $c(x, \widetilde{D})$ be the total number of occurrences of $x$ in $\widetilde{D}$. The collection of transition probabilities for all $x = p_{j-m+1}p_{j-m+2}\cdots p_j$ can be estimated using the set of all *m*-gram and (*m*+1)-gram counts, that is

$$Pr[p_{j+1} = p \,|\, p_{j-m+1}p_{j-m+2}\cdots p_j] = \frac{c(xp,\widetilde{D})}{c(x,\widetilde{D})}. \tag{2}$$

For example, given a 2-gram $x = p_3p_0$ in Figure 4(a), we can get $c(xp_5, \widetilde{D}) = 1$, $c(x, \widetilde{D}) = 3$, and thus the probability that the next location of $x$ will be $p_5$ is $Pr[p_{j+1} = p_5 \,|\, p_3p_0] = 0.333$.

### 3.4 Differential Privacy

Differential privacy (DP) [16] is a notion of privacy that aims to protect the individuals' sensitive information when datasets are released. Given two neighboring databases $D_1$ and $D_2$ (i.e., $D_1$ and $D_2$ differ in only one record), a randomized algorithm *A* satisfies differential privacy if it bounds the probability difference of obtaining the same answer from $D_1$ and $D_2$. The output returned by *A* should be insensitive to the participation of any particular record, and then the inference capability of an adversary is restricted.

In our setting, each record of a dataset is a trajectory, and we assume a user contributes only one trajectory in a dataset. For two neighboring trajectory datasets $D_1'$ and $D_2'$, $D_1'$ can be obtained by removing/adding a user's trajectory from/into $D_2'$. That is, our method aims at guaranteeing user-level unbounded differential privacy.

**Definition 4** (Differential privacy [16]). A randomized algorithm *A* is said to be differentially private if for any two neighboring databases $D_1$ and $D_2$ ($D_1, D_2 \in \mathbb{D}$), and for any possible anonymized output database $o$ ($o \in Range(A)$),

$$Pr[A(D_1) = o] \leq e^\varepsilon \times Pr[A(D_2) = o], \tag{3}$$

where $\varepsilon$ is the privacy budget which defines the privacy level of the mechanism, $\mathbb{D}$ represents the domain of *A* (i.e., the fixed universe of all possible input databases of *A*), and $Range(A)$ denotes the set of all possible outputs of *A*. It is believed that smaller values of $\varepsilon$ lead to stronger privacy guarantee, while higher values pose a lower level of privacy.

One well established technique to achieve differential privacy is the Laplace mechanism [16], which adds appropriate Laplace noise to the real outputs. Let *F* be a function that maps a dataset $\widetilde{D}$ ($\widetilde{D} \in \mathbb{D}$) to a vector of *d* reals, i.e., $F: \mathbb{D} \rightarrow R^d$. The magnitude of the Laplace noise added to $F(\widetilde{D})$ relies on the global sensitivity of *F*.

**Definition 5** (Global sensitivity [16]). For any function $F: \mathbb{D} \rightarrow R^d$, the sensitivity of *F* is

$$\triangle F = \max_{D_1, D_2} \|F(D_1) - F(D_2)\|_1, \tag{4}$$

for any neighboring databases $D_1$ and $D_2$.

**Theorem 1** (Laplace mechanism [16]). *For any function $F: \mathbb{D} \to R^d$, the randomized algorithm $A$ that returns:*

$$A(\widetilde{D}) = F(\widetilde{D}) + Lap\,(\Delta F / \varepsilon), \tag{5}$$

*satisfies $\varepsilon$-differential privacy, where $Lap(\mu)$ is a Laplace random variable with probability density function* $Pr(x|\mu) = \frac{1}{2\mu} e^{\frac{-|x|}{\mu}}$.

Differential privacy has two important properties which we exploit in implementing our method:

**Theorem 2** (Sequential composition [40]). *Let $A_i$ be a randomized algorithm which provides $\varepsilon_i$-differential privacy. Then a sequence of $A_i\,(\widetilde{D})$ ($\widetilde{D} \in \mathbb{D}$) over the database $\widetilde{D}$ provides $(\sum_i \varepsilon_i)$-differential privacy.*

**Theorem 3** (Post-processing [41]): *Let $A_i$ be a randomized algorithm which provides $\varepsilon_i$-differential privacy. Then publicly releasing the output of $A_i\,(\widetilde{D})$ ($\widetilde{D} \in \mathbb{D}$) or using it as an input to another algorithm does not violate $\varepsilon_i$-differential privacy.*

We use these two properties in Section 4 to prove that our noisy prefix tree construction satisfies $\varepsilon_p$-DP, and the noisy $m$-order Markov process construction satisfies $\varepsilon_m$-DP, where $\varepsilon_p$ and $\varepsilon_m$ are the privacy budgets allocated to these two steps. Then, the whole synthetic trajectory generation method DPTraj-PM satisfies $\varepsilon(\varepsilon = \varepsilon_p + \varepsilon_m)$-DP.

## 4 OUR ALGORITHM

### 4.1 Overview

The goal of DPTraj-PM can be stated as follows: Given a raw trajectory dataset $D$, and the differential privacy budget $\varepsilon$, we want to generate a synthetic trajectory dataset $D_{syn}$ such that $\varepsilon$-DP is satisfied and the desirable data utility is retained.

To achieve this goal, we design DPTraj-PM as shown in Figure 5. We discretize all the trajectory points into neighboring cells, model them using a height-$(m + 2)$ prefix tree and an $m$-order Markov process. After adding noise to the model based on DP, we synthesize trajectories from the noisy model. DPTraj-PM consists of three core components: space discretization, private synopsis, and synthetic trajectory generation.

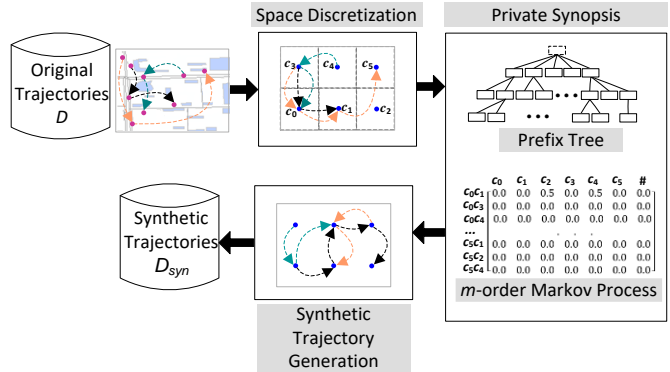

Figure 5: Overview of DPTraj-PM.

### 4.2 Space Discretization

Given a raw trajectory dataset $D = \{T^i| i = 0,1,\ldots, |D|-1\}$ and a continuous spatial domain $Z$ that can cover the whole space of $D$ (e.g., $Z$ is bigger than or equal to the minimum encasing rectangle of $D$'s location points), we first use a reference system $RS$ to discretize $Z$, and map the location points of the trajectories to the anchor points. The reference system is constructed by imposing a $u_h \times u_w$ uniform grid over the space, and choosing the centroids of the cells as anchor points. Let $AP = \{c_i | i = 0, 1, \ldots, u_h \times u_w - 1\}$ denote the anchor point set of $RS$. We summarize the paper's notations in Table 1.

Table 1: Notation summary.

| Symbol | Description | Symbol | Description |
| --- | --- | --- | --- |
| $D$ | A raw trajectory dataset | $neighbor(c_i)$ | The anchor point set of the neighboring cells of $c_i$ together with the stopping symbol # |
| $D_{syn}$ | A synthetic trajectory dataset | $\overline{PT}$ | The noisy prefix tree of $D_c$ |
| $D_c$ | The calibrated dataset of $D$ | $tr(D_c, \vec{v})$ | The set of trajectories in $D_c$ having $prefix(\vec{v}, \overline{PT})$ |
| $h$ | The height of the prefix tree | $|tr(D_c, \vec{v})|$ | The number of trajectories in $tr(D_c, \vec{v})$ |
| $m$ | The order the Markov process | $\Delta tr(\cdot)$ | The sensitivity of $tr(\cdot)$ |
| $\varepsilon$ | The whole privacy budget | $c(\vec{v})$ | A noisy version of $|tr(D_c, \vec{v})|$ |
| $\varepsilon_p$ | The privacy budget allocated to the noisy prefix tree construction | $g$ | A parameter for adjusting the value of $\varepsilon_p$ and $\varepsilon_m$ |
| $\varepsilon_m$ | The privacy budget allocated to the noisy $m$-order Markov process construction | $Q$ | The transition matrix of the $m$-order Markov process |
| $T^i$ | The $i$-th raw trajectory of $D$ | $FM$ | The frequency matrix $FM$ of the $m$-order Markov process |
| $T_c^i$ | The calibrated trajectory of $T^i$ | $F_r$ | The number of rows of $FM$ |
| $Z$ | A spatial domain that can cover the whole space of $D$ | $r_i$ | The label of the $i$-th row of $FM$ |
| $RS$ | A reference system | $n_j$ | The label of the $j$-th column of $FM$ |
| $u_h \times u_w$ | The size of the uniform grid over $Z$ | $E_m = e_0 e_1 \cdots e_{m-1}$ | an *eligible m-gram* |
| $AP$ | The anchor point set of $RS$ | $\psi(\cdot)$ | A query over $D_c$ in the noisy $m$-order Markov process construction |
| $c_i$ | An anchor point of $AP$ | $\Delta\psi(\cdot)$ | The sensitivity of $\psi(\cdot)$ |

If we directly use the transitions between anchor points to simulate the moves between the location points of $D$, each anchor point $c_i \in AP$ has $u_h \times u_w$ transition possibilities, i.e., the $u_h \times u_w - 1$ transitions from $c_i$ to other anchor points of $AP$ and the transition from $c_i$ to the stopping symbol #. This simulation may not be desirable, because some transitions may not be very realistic, e.g., the transition from the bottom-left cell to the top-right cell over a very big region. Improper simulation will not only increase the model size which we will build in Section 4.3, but also introduce unnecessary noise to the data.

To further limit the model size and simulate the connectivity between adjacent location points, we insert points to the raw trajectories by interpolation to make sure that each anchor point $c_i \in AP$ only has at most 9 transition possibilities, i.e., the transitions from $c_i$ to its 8 adjacent cells' anchor points (Figure 6), and the transition from $c_i$ to the stopping symbol #. We call $c_i$'s 8 adjacent cells the neighboring cells of $c_i$. Let *neighbor*$(c_i)$ denote the anchor point set of the neighboring cells of $c_i$ together with the stopping symbol #.

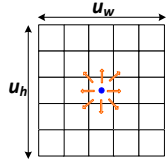

Figure 6: The neighboring cells of an anchor point.

For the sake of simplicity, we use one anchor point to represent the neighboring location points which map to the same anchor point. Then, to indicate the end of a trajectory, a stopping symbol # is inserted into each trajectory as its last element. After this step, $T^i$ becomes $T_c^i = \{p_j^i | j = 0, 1, \ldots, |T_c^i| - 1, p_j^i \in \{AP \cup \{\#\}\}\}$, where $p_j^i$ is the $j$-th anchor point of $T_c^i$ or #, and $|T_c^i|$ denotes the length of $T_c^i$. The raw trajectory dataset $D$ becomes $D_c = \{T_c^i | i = 0, 1, \ldots, |D_c|-1\}$ ($|D_c| = |D|$).

Figure 7 shows an example of space discretization. The reference system in Figure 7(a) has an anchor point set $\{c_0, c_1, c_2, c_3, c_4, c_5\}$, and Figure 7(c) shows the calibrated version of the raw trajectory dataset in Figure 7(b).

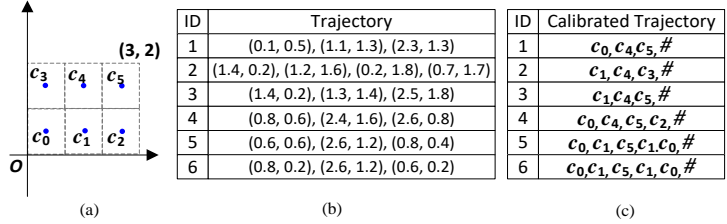

(a)  (b)  (c)

Figure 7: An example of space discretization: (a) a reference system, (b) a raw trajectory dataset, and (c) the calibrated version of the trajectory dataset in Figure 7(b).

### 4.3 Private Synopsis

In this step, we build a mobility model which uses a height-$(m + 2)$ prefix tree to simulate the initial trajectory segments, and an $m$-order Markov process to imitate the next location point picking mechanism, and add noise to this model based on DP. The privacy budgets $\varepsilon_p$ and $\varepsilon_m$ share the whole privacy budget $\varepsilon$ based on a parameter $g$, where $\varepsilon_p = g\varepsilon$, $\varepsilon_m = (1 - g)\varepsilon$, and $0 < g < 1$.

*4.3.1 Noisy Prefix Tree Construction.*

Let $\overline{PT} = (\overline{V}, \overline{E}, \overline{Root})$ be the noisy prefix tree of $D_c$. To satisfy differential privacy, we need to ensure that every possible sequence that can be derived from the anchor point universe $AP$ appears in the noisy prefix tree $\overline{PT}$. Hence, all anchor points of $AP$ appear at the first level. For each node $\bar{v} \in \overline{V}$ at the following levels, its neighboring anchor points and the stopping symbol #, i.e., $neighbor(\bar{v})$, are used to derive its children nodes.

In this process, we assume that the nodes at the lower levels are more important than those at the higher levels in terms of trajectory direction determination. Therefore, a decremental privacy budget allocation approach is exploited.

For each node $\bar{v} \in \overline{V}$, we add Laplace noise to $|tr(D_c, \bar{v})|$ (i.e., the number of sequences in $D_c$ having $prefix(\bar{v}, \overline{PT})$) to guarantee differential privacy. Due to noise addition, each node's noisy count may be not equal to the sum of the noisy counts of its children nodes. To get a consistent and meaningful release, we enforce consistency constraints on the noisy prefix tree $\overline{PT}$ to make sure that for each node $\bar{v} \in \overline{V}$, we have

$$c(\bar{v}) = \sum_{u \in children(\bar{v})} c(u), \tag{6}$$

where $children(\bar{v})$ denotes the children node set of $\bar{v}$.

Let $h$ ($h = m + 2$) be the height of the prefix tree. Given the calibrated trajectory dataset $D_c$, and the anchor point set $AP$, we build the noisy prefix tree $\overline{PT}$ using the following 5 steps.

Step 1: Create an empty tree $\overline{PT} = (\bar{V}, \bar{E}, \overline{Root})$ with a virtual root $\overline{Root}$.

Step 2: Generate the nodes for all anchor points of $AP$ at the first level.

Step 3: For each node $\bar{v} \in \bar{V}$ at the current level $i$ ($i = 1, 2, \ldots, h-1$), calculate $c(\bar{v})$ by adding Laplace noise to $|tr(D_c, \bar{v})|$,

$$c(\bar{v}) = |tr(D_c, \bar{v})| + Lap(\Delta tr(\cdot)/\varepsilon_{p,i}), \tag{7}$$

where

$$\varepsilon_{p,i} = \frac{\log(h-i+\delta)}{\sum_{k=1}^{h-1} \log(h-k+\delta)} \times \varepsilon_p, \tag{8}$$

$\delta$ is an adjustable parameter, and $\Delta tr(\cdot)$ is the sensitivity of $tr(\cdot)$. $|tr(D_c, \bar{v})|$ is regarded as a query over $D_c$ in this phase. For node $\bar{v}$, $tr(\cdot)$ is $|tr(D_c, \bar{v})|$. Since the value of $|tr(\cdot)|$ changes at most one if a trajectory is added to or removed from $D_c$, we have $\Delta tr(\cdot) = 1$.

If $c(\bar{v})$ is smaller than 1 or its $prefix(\bar{v}, \overline{PT})$ ends with the stopping symbol #, the node will not be expanded further; otherwise, generate its children nodes for all elements of $neighbor(\bar{v})$ at the following level.

Step 4: Repeat Step 3 until the maximum tree height $h$ is reached.

Step 5: Enforce consistency constraints across the levels of $\overline{PT}$ from top to bottom. For each node $\bar{v} \in \bar{V}$ at level $i$ ($i = 1, 2, \ldots, h-1$), update $c(\bar{v})$ by

$$c(\bar{v}) = \frac{c(\bar{v})}{\sum_{u \in children(par(\bar{v}))} c(u)} \times c(par(\bar{v})), \tag{9}$$

where $par(\bar{v})$ is the parent node of $\bar{v}$. In our method, $c(\overline{Root}) = |D_c|$.

**Theorem 4.** *Noisy prefix tree construction satisfies $\varepsilon_p$-DP.*

*Proof.* The actual input of the noisy prefix tree construction is the raw trajectory dataset $D$. According to the space discretization in Section 4.2, we know that for two neighboring datasets $D$ and $D'$ whose spatial domains are both $Z$, their corresponding calibrated versions $D_c$ and $D_c'$ are also two neighboring datasets. Therefore, we just need to prove that the noisy prefix tree construction satisfies $\varepsilon_p$-DP by regarding $D_c$ and $D_c'$ as the inputs.

The noisy prefix tree construction contains two parts: noise addition (Step 1- Step 4) and consistency constraints enforcement (Step 5). Since the consistency constraint enforcement does not get access to $D_c$, it is a postprocessing step and consumes zero privacy budget. Then, we just need to prove the noise addition consumes a privacy budget $\varepsilon_p$. That is, given two neighboring datasets $D_c$ and $D_c'$, we need to prove $\frac{Pr(A_p(D_c)=\overline{PT})}{Pr(A_p(D_c')=\overline{PT})} \leq e^{\varepsilon_p}$, where $A_p$ denotes the noise addition.

Let $\varepsilon_{\bar{v}}$ be a node $\bar{v}$'s ($\bar{v} \in \bar{V}$) privacy budget. We have

$$\frac{Pr(A_p(D_c)=\overline{PT})}{Pr(A_p(D_c')=\overline{PT})} = \prod_{\bar{v} \in \bar{V}} \frac{\exp\left(-\varepsilon_{\bar{v}} \frac{|c(\bar{v})-|tr(D_c,\bar{v})||}{\Delta tr(\cdot)}\right)}{\exp\left(-\varepsilon_{\bar{v}} \frac{|c(\bar{v})-|tr(D_c',\bar{v})||}{\Delta tr(\cdot)}\right)} \leq \prod_{\bar{v} \in \bar{V}} \exp\left(\frac{\varepsilon_{\bar{v}}||tr(D_c,\bar{v})|-|tr(D_c',\bar{v})||}{\Delta tr(\cdot)}\right). \tag{10}$$

Note that a trajectory can only affect one root-to-leaf path of $\overline{PT}$. Let $\eta$ be the set of all nodes along such a path. Then, Equation (10) can be rewritten as

$$\frac{Pr(A_p(D_c)=\overline{PT})}{Pr(A_p(D_c')=\overline{PT})} \leq \prod_{\bar{v} \in \bar{V}} \exp\left(\frac{\varepsilon_{\bar{v}}||tr(D_c,\bar{v})|-|tr(D_c',\bar{v})||}{\Delta tr(\cdot)}\right) = \prod_{\bar{v} \in \eta} \exp(\varepsilon_{\bar{v}}) = \exp(\sum_{\bar{v} \in \eta} \varepsilon_{\bar{v}}). \tag{11}$$

Since $\sum_{\bar{v} \in \eta} \varepsilon_{\bar{v}} \leq \varepsilon_p$, we get

$$\frac{Pr(A_p(D_c)=\overline{PT})}{Pr(A_p(D_{c'})=\overline{PT})} \leq \exp(\sum_{\bar{v} \in \eta} \varepsilon_{\bar{v}}) \leq e^{\varepsilon_p}. \tag{12}$$

That is, the noise addition satisfies $\varepsilon_p$-DP, and thus the noisy prefix tree construction satisfies $\varepsilon_p$-DP.

*4.3.2 Noisy m-order Markov Process Construction.*

We use a transition matrix $Q$ to represent an $m$-order Markov process in this step. We first calculate a frequency matrix $FM$ by scanning all the trajectories of $D_c$, add noise to $FM$, and then derive the transition matrix $Q$ from the noisy $FM$ (denoted by $\overline{FM}$).

To satisfy differential privacy, we need to guarantee that the frequency of every possible $(m + 1)$-gram that can be derived from the universe $AP \cup \{\#\}$ appears in $\overline{FM}$. For each trajectory of $D_c$, the elements are adjacent cells' anchor points. Therefore, we just need to consider *eligible* $(m + 1)$-grams. We call an $(m+1)$-gram $E_{m+1} = e_0 e_1 \cdots e_m$ $(e_i \in \{AP \cup \{\#\}\}$, $i = 0, 1, \ldots, m$) an *eligible* $(m + 1)$-gram if for $\forall 0 \leq i < m$, we have $e_{i+1} \in neighbor(e_i)$. $E_{m+1}$ can be derived from an *eligible* $m$-gram $E_m = e_0 e_1 \cdots e_{m-1}$ by $E_{m+1} = E_m e_m$.

Let $F_r$ be the number of *eligible* $m$-grams of $D_c$. We construct a frequency matrix $FM$ that has $F_r$ rows and $|AP| + 1$ columns, with $FM_{i,j}$ representing the element on the $i$-th row and the $j$-th column of $FM$. For ease of description, we put a label $r_i$ in front of the $i$-th $(i = 0, 1, \ldots, F_r - 1)$ row of $FM$ (e.g., $C_0 C_1$ in front of the first row of the frequency matrix in Figure 8(a)), and a label $n_j$ in front of the $j$-th $(j = 0, 1, \ldots, |AP|)$ column of $FM$ (e.g., $C_0$ in front of the first column of the frequency matrix in Figure 8(a)). Here, $r_i$ is an *eligible* $m$-gram composed of adjacent cells' anchor points, and $n_j$ is an anchor point. Then, $FM_{i,j}$ stores the frequency of the $(m + 1)$-gram $r_i n_j$. For the $i$-th row of $FM$, only the frequencies of the *eligible* $(m+1)$-grams that can be derived from $r_i$ are calculated.

|       | $C_0$ | $C_1$ | $C_2$ | $C_3$ | $C_4$ | $C_5$ | #    |
|-------|-------|-------|-------|-------|-------|-------|------|
| $C_0C_1$ | 0.0   | 0.0   | 0.0   | 0.0   | 0.0   | 0.50  | 0.0  |
| $C_0C_3$ | 0.0   | 0.0   | 0.0   | 0.0   | 0.0   | 0.00  | 0.0  |
| $C_0C_4$ | 0.0   | 0.0   | 0.0   | 0.0   | 0.0   | 0.83  | 0.0  |
| ...   |       |       |       |       |       |       |      |
| $C_5C_1$ | 0.5   | 0.0   | 0.0   | 0.0   | 0.0   | 0.0   | 0.00 |
| $C_5C_2$ | 0.0   | 0.0   | 0.0   | 0.0   | 0.0   | 0.0   | 0.33 |
| $C_5C_4$ | 0.0   | 0.0   | 0.0   | 0.0   | 0.0   | 0.0   | 0.00 |

(a)

|       | $C_0$ | $C_1$ | $C_2$ | $C_3$ | $C_4$ | $C_5$ | #    |
|-------|-------|-------|-------|-------|-------|-------|------|
| $C_0C_1$ | 0.1   | 0.0   | 1.2   | 0.0   | 0.0   | 0.50  | 0.2  |
| $C_0C_3$ | 0.2   | 0.1   | 0.0   | 0.0   | 0.2   | 0.00  | 0.0  |
| $C_0C_4$ | 1.0   | 0.0   | 0.0   | 0.2   | 0.0   | 0.60  | 0.2  |
| ...   |       |       |       |       |       |       |      |
| $C_5C_1$ | 0.2   | 0.0   | 0.1   | 0.0   | 0.0   | 0.0   | 0.2  |
| $C_5C_2$ | 0.0   | 0.1   | 0.0   | 0.0   | 0.2   | 0.2   | 0.5  |
| $C_5C_4$ | 0.2   | 0.3   | 0.0   | 0.0   | 0.0   | 0.0   | 0.1  |

(b)

|       | $C_0$ | $C_1$ | $C_2$ | $C_3$ | $C_4$ | $C_5$ | #    |
|-------|-------|-------|-------|-------|-------|-------|------|
| $C_0C_1$ | 0.05  | 0.0   | 0.6   | 0.0   | 0.0   | 0.25  | 0.10 |
| $C_0C_3$ | 0.40  | 0.2   | 0.0   | 0.0   | 0.4   | 0.00  | 0.00 |
| $C_0C_4$ | 0.50  | 0.0   | 0.0   | 0.1   | 0.0   | 0.30  | 0.10 |
| ...   |       |       |       |       |       |       |      |
| $C_5C_1$ | 0.40  | 0.0   | 0.2   | 0.0   | 0.0   | 0.0   | 0.40 |
| $C_5C_2$ | 0.00  | 0.1   | 0.0   | 0.0   | 0.2   | 0.2   | 0.50 |
| $C_5C_4$ | 0.33  | 0.5   | 0.0   | 0.0   | 0.0   | 0.0   | 0.17 |

(c)

Figure 8: (a) The frequency matrix $FM$ of the calibrated trajectory dataset in Figure 7(c), (b) an example of the noisy frequency matrix $\overline{FM}$ of Figure 8(a), and (c) the transition matrix $Q$ derived from Figure 8(b).

We add Laplace noise to the elements of $FM$ to guarantee differential privacy, and derive the transition matrix $Q$ from the noisy $FM$, i.e., $\overline{FM}$, by the following 4 steps:

Step 1: Initialize two matrices $FM = \{FM_{i,j} = 0 | i = 0, 1, \ldots, F_r - 1, j = 0, 1, \ldots, |AP|\}$ and $\overline{FM} = \{\overline{FM}_{i,j} = 0 | i = 0, 1, \ldots, F_r - 1, j = 0, 1, \ldots, |AP|\}$.

Step 2: Scan each trajectory $T_c^i$ ($i = 0, 1, \ldots, |D_c|-1$) of $D_c$ to get the frequency of every $(m+1)$-gram in it. For the *eligible* $(m + 1)$-gram $r_i n_j$, we calculate its frequency $FM_{i,j}$ by a query $\psi(\cdot)$ over $D_c$,

$$FM_{i,j} = \psi(r_i n_j, D_c) = \sum_{T_c^i \in D_c} \frac{fc(r_i n_j, T_c^i)}{|T_c^i| - m}, \tag{13}$$

where $fc(r_i n_j, T_c^i)$ is the number of $r_i n_j$ in $T_c^i$. The query $\psi(\cdot)$ is the function we want to make differentially private in this phase.

Step 3: For each element $\overline{FM}_{i,k} \in \overline{FM}$ ($k$ satisfies that $r_i n_k$ is an *eligible* ($m + 1$)-gram) of $\overline{FM}$, calculate its value by adding Laplace noise to $FM_{i,k}$,

$$\overline{FM}_{i,k} = FM_{i,k} + Lap(\Delta\psi(\cdot)/\varepsilon_m), \tag{14}$$

where $\Delta\psi(\cdot)$ is the sensitivity of $\psi(\cdot)$. Since the value of $|\psi(\cdot)|$ changes at most one if a trajectory is added to or removed from $D_c$, we have $\Delta\psi(\cdot) = 1$.

Step 4: Initialize a transition matrix $Q = \{Q_{i,j} = 0 | i = 0, 1, \ldots, F_r - 1, j = 0, 1, \ldots, |AP|\}$ that has the same size as $FM$, and compute $Q_{i,k} \in Q$ by

$$Q_{i,k} = \overline{FM}_{i,k} / \sum_{u=0}^{|AP|} \overline{FM}_{i,u}. \tag{15}$$

An example of the 2-order Markov process construction for the calibrated trajectory dataset in Figure 7(c) is shown in Figure 8. For simplicity, six rows of the frequency matrix $FM$ have been listed in Figure 8(a). After adding noise, an example of the noisy frequency matrix $\overline{FM}$ of Figure 8(a) is shown in Figure 8(b). Figure 8(c) gives the transition matrix $Q$ derived from Figure 8(b).

**Theorem 5.** *Noisy m-order Markov process construction satisfies $\varepsilon_m$-DP.*

*Proof.* The noisy *m*-order Markov process construction has two parts: noisy frequency matrix construction (Step 1- Step 3) and transition matrix construction (Step 4). Since the transition matrix construction is a postprocessing step, which consumes zero privacy budget, we just need to prove the noisy frequency matrix construction consumes a privacy budget $\varepsilon_m$. In other words, given two neighboring datasets $D_c$ and $D_c'$, we need to prove $\frac{Pr(A_m(D_c)=\overline{FM})}{Pr(A_m(D_{c'})=\overline{FM})} \leq e^{\varepsilon_m}$, where $A_m$ represents the noisy frequency matrix construction. A trajectory $T_c^i (i = 0,1,\ldots, |D_c|-1)$ of $D_c$ has at most $(|T_c^i| - m)$ different kinds of ($m + 1$)-grams, and hence can only affect at most $(|T_c^i| - m)$ elements of $FM$. Let $\beta$ be the set of all elements of $FM$ that are affected by $T_c^i$, and $FM'$ denote the frequency matrix of $D_c'$. We have

$$\frac{Pr(A_m(D_c)=\overline{FM})}{Pr(A_m(D_{c'})=\overline{FM})} = \prod_{FM_{i,j}\in FM} \frac{\exp\left(-\varepsilon_m \frac{|\overline{FM}_{i,j}-FM_{i,j}|}{\Delta\psi(\cdot)}\right)}{\exp\left(-\varepsilon_m \frac{|\overline{FM}_{i,j}-FM_{i,j}'|}{\Delta\psi(\cdot)}\right)} \leq \prod_{FM_{i,j}\in FM} \exp\left(\frac{\varepsilon_m |FM_{i,j}-FM_{i,j}'|}{\Delta\psi(\cdot)}\right)$$

$$= \prod_{FM_{i,j}\in FM} \exp(\varepsilon_m |FM_{i,j} - FM_{i,j}'|) = \prod_{FM_{i,j}\in \beta} \exp(\varepsilon_m |FM_{i,j} - FM_{i,j}'|)$$

$$= \exp(\varepsilon_m \sum_{FM_{i,j}\in \beta} |FM_{i,j} - FM_{i,j}'|). \tag{16}$$

According to Equation (13), we can get $\sum_{FM_{i,j}\in \beta} |FM_{i,j} - FM_{i,j}'| = 1$. Then, Equation (16) can be rewritten as

$$\frac{Pr(A_m(D_c)=\overline{FM})}{Pr(A_m(D_{c'})=\overline{FM})} \leq \exp(\varepsilon_m \sum_{FM_{i,j}\in \beta} |FM_{i,j} - FM_{i,j}'|) = e^{\varepsilon_m}. \tag{17}$$

That is, the noisy frequency matrix construction satisfies $\varepsilon_m$-DP, and thus the noisy *m*-order Markov process construction satisfies $\varepsilon_m$-DP.

## 4.4 Synthetic Trajectory Generation

In this step, we use the noisy prefix tree $\overline{PT}$ and the transition matrix $Q$ to generate the synthetic trajectory dataset $D_{syn}$. $\overline{PT}$ is employed to determine the initial trajectory segment, and $Q$ is used to pick the next anchor point for each trajectory.

The nodes of $\overline{PT}$ can be divided into two types: non-generalized node and generalized node. A non-generalized node associates with an anchor point sequence that ends with the stopping symbol #, while a generalized node does not.

For a non-generalized node $v_a$ of $\overline{PT}$, we generate its corresponding trajectories by appending $c(v_a)$ copies of $prefix(v_a, \overline{PT})$ to the output.

For a generalized node $v_b$ at the last level (i.e., the $(h-1)$-th level) of $\overline{PT}$, $c(v_b)$ trajectories with the initial segment $prefix(v_b, \overline{PT})$ are generated. Let $T_c = prefix(v_b, \overline{PT}) = e_0 \cdots e_{h-m-1} e_{h-m} \cdots e_{h-2}$ ($e_{h-2} \neq$ #) be one of such trajectories, and $r_u = e_{h-m-1} e_{h-m} \cdots e_{h-2}$ be the last $m$ anchor point sequence of $T_c$, which is an *eligible* $m$-gram. We select an anchor point $n_t$ ($n_t \in \{AP \cup \{\#\}\}$) as the next point of $T_c$ with a probability $Q_{u,t}$ ($Q_{u,t} \in Q$), and append it to $T_c$. This anchor point selection and appending process is repeated until the stopping symbol # is appended to $T_c$ or the maximum length $l_{max}$ is reached.

**Theorem 6.** *DPTraj-PM satisfies $\varepsilon$-DP.*

*Proof.* DPTraj-PM contains 3 parts: space discretization, private synopsis, and synthetic trajectory generation. The space discretization and the synthetic trajectory generation which performs a post-processing step do not consume any privacy budget. The private synopsis is composed of two parts: the noisy prefix tree construction and the $m$-order Markov process construction, whose inputs are both $D_c$. According to the sequential composition property, we get that the privacy budget of the private synopsis is $\varepsilon = \varepsilon_p + \varepsilon_m$. Therefore, DPTraj-PM satisfies $\varepsilon$-DP.

## 5 EXPERIMENTAL RESULTS AND ANALYSIS

### 5.1 Experiment Setup

#### 5.1.1 Datasets.

We perform our experiments over two real-world datasets: Taxi [24] and Geolife [42].
- **Taxi.** Taxi is the dataset of the Taxi Service Prediction Challenge at ECML-PKDD 2015. It consists of a complete year (from 01/07/2013 to 30/06/2014) data of traces for 442 taxis running in Porto. We extract two datasets (namely Taxi-1 and Taxi-2) from the first 904729 traces of Taxi using two bounding boxes. The bounding box, which is defined by the lower-left corner and the upper-right corner points of a rectangle, is utilized as the spatial domain $Z$ in DPTraj-PM.
- **Geolife.** Geolife was collected in the Geolife project of Microsoft Research Asia. It contains 17621 GPS trajectories of 182 users from April 2007 to August 2012 in Beijing. We extract one dataset from Geolife. In the experiments, we assume all the trajectories of Geolife belong to different users.

The trajectory datasets are summarized in Table 2. To clearly illustrate the location point density distributions of the raw datasets in different regions, we respectively apply a 20×20 uniform grid on Taxi-1 and Taxi-2, and 6×6 uniform grid on Geolife, and calculate the number of trajectories passing through each cell for the heat maps in Figure 9. The average location points per cell is shown in the last column of Table 2.

Table 2: Datasets summary.

| Dataset | Bounding box | Area | $|D|$ | Trajectory density ($|D|$ / Area) | Average location points per cell |
|---|---|---|---|---|---|
| Taxi-1 | (41.104N, 8.665W), (41.250N, 8.528W) | 11.49km×16.23km | 893067 | 4789.007 | 18393.68 |
| Taxi-2 | (41.064N, 8.662W), (41.210N, 8.525W) | 11.49km×16.23km | 889515 | 4769.960 | 17812.71 |
| Geolife | (39.788N,116.148E), (40.093N, 116.612E) | 39.69km×33.95km | 17073 | 12.670 | 1231.528 |

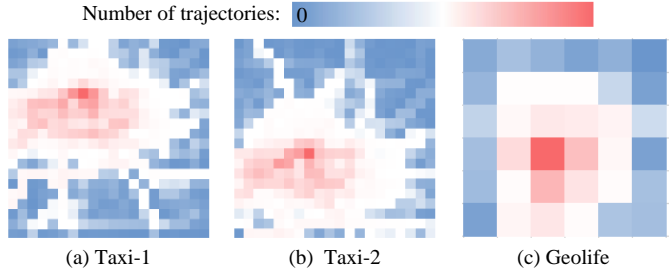

(a) Taxi-1    (b) Taxi-2    (c) Geolife

Figure 9: Location point density distributions of the raw datasets: (a) Taxi-1, (b) Taxi-2, and (c) Geolife.

### 5.1.2 Utility Metrics.

To evaluate the utility of the synthetic database $D_{syn}$, we impose a uniform grid $U$ over the spatial domain of $D$, and calculate seven metrics at the cell level. We regard each cell as a discrete location. Let $Cell = \{cell_i | i = 0, 1, ..., \alpha - 1\}$ be the cell set of $U$, and $\alpha$ be the number of $Cell$'s elements.

- **Location Metrics.** The similarities and discrepancies between locations' popularity are important for many geodata analysis tasks, such as hotpot discovery and POI recommendation. We use three metrics to evaluate the quality of similarities and discrepancies between locations' popularity: location visit average relative error (Location *AvRE*), location visit proportion, and location Kendall-tau (Location *KT*) coefficient [20-21].

    Let $pop(cell_i, D)$ be the number of times $cell_i$ visited by the traces of $D$. We define the visit *relative error* (*RE*) of $cell_i$ as

$$RE = \frac{|pop(cell_i, D) - pop(cell_i, D_{syn})|}{max\{pop(cell_i, D), \lambda\}},$$

where $\lambda$ is a *sanity bound* that mitigates the effect of cells with extremely small visits. The Location *AvRE* is calculated by averaging all the *RE* of all the locations of *Cell*. In the experiments, we set $\lambda = 0.1\% \times |D|$.

The location visit proportion can be used to assess the geographic and semantic similarity between locations in $D$ and $D_{syn}$ [7]. Let $Cell_U^n = \{cell_i | pop(cell_i, D) \geq pop(cell_{i+1}, D), cell_i \in Cell, i = 0, 1, ..., n-1, n \leq \alpha\}$ be the top-$n$ visited cells in the region of $D$. We define the visit proportion $prop(cell_i, D)$ for a $cell_i \in Cell_U^n$ as

$$prop(cell_i, D) = \frac{pop(cell_i, D)}{\sum_{cell_j \in Cell} pop(cell_j, D)}.$$

We set $n = 20$ in the experiments.

We use the location *KT* coefficient [20-21] to measure the similarities and discrepancies between locations' popularity ranking in $D$ and $D_{syn}$. The pair $(cell_i, cell_j)$ is concordant if the popularity ranks of $cell_i$ and $cell_j$ in sorted order agree in $D$ and $D_{syn}$. That is, a cell pair is said to be concordant if one of the following conditions is satisfied:

$$(pop(cell_i, D) > pop(cell_j, D)) \wedge (pop(cell_i, D_{syn}) > pop(cell_j, D_{syn}))$$
$$(pop(cell_i, D) < pop(cell_j, D)) \wedge (pop(cell_i, D_{syn}) < pop(cell_j, D_{syn}))$$.

Then, the location *KT* coefficient [20-21] is calculated by

$$\text{Location } KT = \frac{(\text{\#of concordant location pairs}) - (\text{\#of discordant location pairs})}{\alpha(\alpha-1)/2}.$$

- **Frequent Pattern Metrics. Mining.** Frequent patterns is a central task in many applications, including traffic flow analysis and route navigation. We use the frequent pattern average relative error (FP *AvRE*) [17, 20] and frequent pattern Kendall-tau (FP *KT*) coefficient [21, 23] to measure whether the FPs are well preserved in $D_{syn}$.

  Assume a pattern $P$ is represented by an ordered list of the cells of *Cell*. Let $sup(P, D)$ denote the number of occurrences of $P$ in $D$, and $\text{FP}_U^k(D) = \{P_i | i = 0, 1, \ldots, k-1\}$ be the top-$k$ patterns in $D$. The FP *AvRE* [17, 20] is computed by

  $$\text{FP } AvRE = \frac{\sum_{P_i \in \text{FP}_U^k(D)} \frac{|sup(P_i, D) - sup(P_i, D_{syn})|}{sup(P_i, D)}}{k}.$$

  The FP *KT* coefficient [21, 23] is used to evaluate the similarities and discrepancies between frequent patterns' popularity ranking in $D$ and $D_{syn}$. For two frequent patterns $P_i, P_j \in \text{FP}_U^k(D)$ ($i, j = 0, 1, \ldots, k-1$) of $\text{FP}_U^k(D)$, the pair $(P_i, P_j)$ is concordant if one of the following conditions holds:

  $$(sup(P_i, D) > sup(P_j, D)) \wedge (sup(P_i, D_{syn}) > sup(P_j, D_{syn}))$$
  $$(sup(P_i, D) < sup(P_j, D)) \wedge (sup(P_i, D_{syn}) < sup(P_j, D_{syn}))$$

  Accordingly, the FP *KT* coefficient can then be calculated by

  $$\text{FP } KT = \frac{(\text{\#of concordant FP pairs}) - (\text{\#of discordant FP pairs})}{k(k-1)/2}.$$

  In our experiments, the parameter $k = 200$, the length of each pattern is from 2 to 8.

- **Trip Metrics.** Trip information is also very helpful in many fields, such as urban planning and transportation research. We use two metrics for trip analysis: trip error [20-21, 23] and length error [20-21].

  The trip error [20-21, 23] is used to measure how well the correlations between the starting and ending regions of the original traces are preserved. Let $trip(U, D)$ be the empirical trip distribution of $D$, i.e., the distribution of all possible pairs of starting and ending cells. The trip error is measured as the Jensen-Shannon divergence between $trip(U, D)$ and $trip(U, D_{syn})$: $JSD(trip(U, D), trip(U, D_{syn}))$.

  The length of a trip $T^i$ is calculated by summing up the distance between adjacent points in $T^i$ [20-21]. Let $Len(D)$ be the empirical distribution of the trip lengths on $D$, where the trip lengths are quantized into 20 equal width buckets. The length error is calculated as: $JSD(Len(D), Len(D_{syn}))$.

  In the calculation of the utility metrics, we impose a 6×6 uniform grid on Geolife, and a 20×20 uniform grid on Taxi-1 and Taxi-2 to get the cell sets, respectively.

## 5.2 Comparison with Related Works

### 5.2.1 Related Works.

DPTraj-PM is compared with four most relevant works: DPT [19], AdaTrace [20], DP-MODR [21] and LDPTrace [22]. The implementations of [19-20,22] were obtained from the respective authors. In the comparison, the parameters recommended by the authors in their paper or their codes were used when applicable (Table 3). For [19], we have tested different resolution sets and adopt the set that produces the best average data utility. [20] uses a uniform grid at the first level and an adaptive grid division at the second level. We have tested different grid sizes for the first level, and use the one that yields the highest average data utility. For [21], we have tested different heights of the cost-sensitive path trees, and utilize the height that produces the best average data utility. [22] uses a uniform grid. Because the Laplace mechanisms and the exponential mechanisms used in the five methods are probabilistic, we repeat each experiment five times and report the average results for [19-22] and DPTraj-PM.

Table 3: Experimental parameter setting.

| Dataset | [19] | [20] | [21] | [22] | Proposed |
|---|---|---|---|---|---|
| Geolife | resolution set = {0.016, 0.032, 0.048, 0.064} percentage budget for model selection: 0.1 geometric_adaptiveprune: 0.1 dir: 1.20:15 | grid size at the first level: 6×6 $\varepsilon_1=0.05\varepsilon$, $\varepsilon_2=0.35\varepsilon$, $\varepsilon_3=0.50\varepsilon$, $\varepsilon_4=0.10\varepsilon$ | grid division: 6×6 $h_{max}=4$ $\varepsilon_1=\varepsilon_2=\varepsilon_3=\varepsilon/3$ | grid division: 6×6 $\varepsilon_1=\varepsilon/10$ $\varepsilon_2+\varepsilon_3=9\varepsilon/10$ | grid division: 6×6 $m=2, g=0.6$ $\delta=0.8$ $l_{max}=6\times6$ |
| Taxi-1 | resolution set = {0.004, 0.005, 0.006} percentage budget for model selection: 0.1 geometric_adaptiveprune: 0.1 dir: 1.20:15 | grid size at the first level: 12×12 $\varepsilon_1=0.05\varepsilon$, $\varepsilon_2=0.35\varepsilon$, $\varepsilon_3=0.50\varepsilon$, $\varepsilon_4=0.10\varepsilon$ | grid division: 20×20 $h_{max}=3$ $\varepsilon_1=\varepsilon_2=\varepsilon_3=\varepsilon/3$ | grid division: 20×20 $\varepsilon_1=\varepsilon/10$ $\varepsilon_2+\varepsilon_3=9\varepsilon/10$ | grid division: 20×20 $m=3, g=0.6$ $\delta=0.8$ $l_{max}=20\times20$ |
| Taxi-2 | resolution set = {0.004, 0.005, 0.006} percentage budget for model selection: 0.1 geometric_adaptiveprune: 0.1 dir: 1.20:15 | grid size at the first level: 12×12 $\varepsilon_1=0.05\varepsilon$, $\varepsilon_2=0.35\varepsilon$, $\varepsilon_3=0.50\varepsilon$, $\varepsilon_4=0.10\varepsilon$ | grid division: 20×20 $h_{max}=3$ $\varepsilon_1=\varepsilon_2=\varepsilon_3=\varepsilon/3$ | grid division: 20×20 $\varepsilon_1=\varepsilon/10$ $\varepsilon_2+\varepsilon_3=9\varepsilon/10$ | grid division: 20×20 $m=3, g=0.6$ $\delta=0.8$ $l_{max}=20\times20$ |

### 5.2.2 Utility Comparisons.

We compare DPTraj-PM with [19-22] using the metrics in Section 5.1.2 with varying privacy budget $\varepsilon$. Table 4 summarizes the results, with the best result in each category being shown in bold. Using the location visit proportion metric, the probability distribution of visited top-$n$ regions in the generated trajectories for each technique has been compared to the true datasets. The results over Taxi-1, Taxi-2 and Geolife datasets are illustrated in Figure 2, Figure 10 and Figure 11, respectively.

We have three general observations. First, DPTraj-PM provides superior utility in general when subjected to the same level of privacy. The improvements are substantial. Although its location visit *AvRE* value are not the smallest on Geolife when $\varepsilon = 0.1$, its performance improves very quickly with the increase of $\varepsilon$. Besides, the difference between its length error value and the best case is no more than 1%. Second, we find larger $\varepsilon$ leads to less errors and better utility in general. This conforms to the theoretical analysis that a higher $\varepsilon$ results in less noise and thus a more accurate result. Third, Geolife which has a lower trajectory density and a lower average location points per cell tends to be more sensitive to the changes of $\varepsilon$ than Taxi-1 and Taxi-2. On the other hand, Taxi-1 which has a higher trajectory density and a higher average location points per cell than Taxi-2 and Geolife does not always preserve more data utility. That's probably because the data utility is also affected by the location point density distribution (Figure 9). [19-22] and DPTraj-PM rely on the connections between cells to model the trajectories. For different datasets, each cell may contain different trajectories, which may result in pretty different aggregate properties, and thus different data utility.

Table 4: Comparing DPTraj-PM with its competitors. Best result in each category is shown in bold. For location *KT* and FP *KT*, higher values are better. For remaining metrics, lower values are better.

| Metrics | $\varepsilon$ | Taxi-1 | | | | | Taxi-2 | | | | | Geolife | | | | |
|---|---|---|---|---|---|---|---|---|---|---|---|---|---|---|---|---|
| | | [19] | [20] | [21] | [22] | DPTraj-PM | [19] | [20] | [21] | [22] | DPTraj-PM | [19] | [20] | [21] | [22] | DPTraj-PM |
| Location Visit *AvRE* | 0.1 | 1.453 | 2.704 | 0.655 | 2.370 | **0.395** | 1.971 | 2.977 | 0.627 | 2.557 | **0.381** | **0.639** | 3.237 | 0.902 | 1.302 | 0.967 |
| | 0.5 | 0.651 | 1.311 | 0.668 | 2.224 | **0.374** | 0.692 | 1.109 | 0.622 | 2.291 | **0.363** | 0.351 | 1.090 | 0.577 | 1.216 | **0.257** |
| | 1 | 0.458 | 1.157 | 0.656 | 2.271 | **0.383** | 0.486 | 0.873 | 0.604 | 1.973 | **0.370** | 0.400 | 0.580 | 0.625 | 1.407 | **0.199** |
| Location *KT* | 0.1 | 0.667 | 0.621 | 0.626 | 0.527 | **0.730** | 0.587 | 0.554 | 0.684 | 0.513 | **0.742** | 0.756 | 0.721 | 0.702 | 0.633 | **0.763** |
| | 0.5 | 0.739 | 0.665 | 0.698 | 0.542 | **0.814** | 0.716 | 0.638 | 0.751 | 0.525 | **0.820** | 0.833 | 0.853 | 0.785 | 0.694 | **0.881** |
| | 1 | 0.771 | 0.673 | 0.702 | 0.533 | **0.845** | 0.757 | 0.657 | 0.750 | 0.565 | **0.842** | 0.832 | 0.882 | 0.793 | 0.660 | **0.904** |
| FP *AvRE* | 0.1 | 0.584 | 0.744 | 1.350 | 0.668 | **0.362** | 0.674 | 0.700 | 1.287 | 0.656 | **0.383** | 0.988 | 0.782 | 5.376 | 0.730 | **0.687** |
| | 0.5 | 0.569 | 0.745 | 1.260 | 0.656 | **0.308** | 0.613 | 0.659 | 1.275 | 0.656 | **0.327** | 0.754 | 0.594 | 1.927 | 0.703 | **0.528** |
| | 1 | 0.582 | 0.730 | 1.255 | 0.658 | **0.293** | 0.607 | 0.654 | 1.271 | 0.650 | **0.323** | 0.787 | 0.525 | 1.405 | 0.715 | **0.470** |
| FP *KT* | 0.1 | 0.372 | 0.098 | 0.217 | 0.423 | **0.549** | 0.264 | 0.150 | 0.255 | 0.431 | **0.478** | 0.313 | 0.189 | -0.013 | 0.359 | **0.485** |
| | 0.5 | 0.421 | 0.012 | 0.185 | 0.425 | **0.599** | 0.273 | 0.158 | 0.217 | 0.437 | **0.507** | 0.212 | 0.262 | 0.079 | 0.335 | **0.558** |
| | 1 | 0.389 | -0.015 | 0.149 | 0.425 | **0.612** | 0.286 | 0.163 | 0.196 | 0.445 | **0.507** | 0.125 | 0.361 | 0.003 | 0.346 | **0.584** |
| Trip Error | 0.1 | 0.302 | 0.437 | 0.486 | 0.249 | **0.106** | 0.316 | 0.380 | 0.492 | 0.257 | **0.126** | 0.138 | 0.303 | 0.305 | 0.124 | **0.086** |
| | 0.5 | 0.260 | 0.436 | 0.487 | 0.246 | **0.093** | 0.252 | 0.354 | 0.499 | 0.256 | **0.110** | 0.116 | 0.165 | 0.234 | 0.125 | **0.045** |
| | 1 | 0.244 | 0.440 | 0.486 | 0.248 | **0.090** | 0.238 | 0.369 | 0.500 | 0.253 | **0.109** | 0.119 | 0.085 | 0.246 | 0.125 | **0.040** |
| Length Error | 0.1 | 0.025 | 0.032 | 0.012 | 0.007 | **0.002** | 0.036 | 0.044 | 0.019 | 0.010 | **0.009** | 0.031 | 0.046 | 0.049 | **0.003** | 0.006 |
| | 0.5 | 0.012 | 0.014 | 0.008 | 0.007 | **0.004** | 0.025 | 0.024 | 0.015 | **0.008** | 0.015 | 0.009 | 0.013 | 0.006 | **0.003** | 0.005 |
| | 1 | 0.007 | 0.012 | 0.008 | 0.007 | **0.004** | 0.018 | 0.023 | 0.016 | **0.007** | 0.016 | 0.025 | 0.007 | **0.002** | 0.004 | 0.007 |

Then, we do an in-depth analysis by studying the results of each method one by one, starting with DPT [19]. DPT chooses the optimum prefix tree set with different resolutions to model the trajectories. It may traverse from one noisy prefix tree to another, or scan the root-to-leaf paths of the same noisy prefix tree many times to generate the synthetic traces. For the former case, the accuracy of the generated trajectory segments may be reduced after noisy prefix tree transfer, because no information of the prior trajectory segment except the last node is used in the determination of the next point. For the latter, some extra noise may be introduced, because the trace has to contain all the noise added along the root-to-leaf paths every time, while only the noise at the last level are necessary for the next point picking. DPTraj-PM uses one prefix tree to simulate the initial trajectory segments, and an *m*-order Markov process for the next point generation. Some unnecessary data utility loss may have been avoided, and thus DPTraj-PM is superior to DPT in general.

The performance of AdaTrace [20] is inferior to DPTraj-PM. The reason may be that its next location point picking approach depends on a 1-order Markov process, which may be not adequate to simulate the individual mobility behaviors. Besides, the 1-order Markov process has to share the whole privacy budget with other 3 core components, which may further limit its data accuracy.

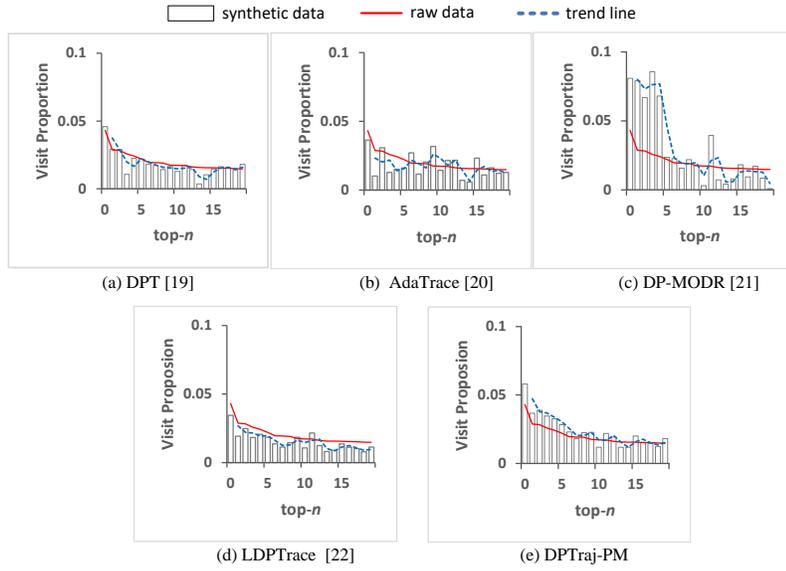

Figure 10: Probability distribution of top-$n$ visited regions for real and synthetic trajectories generated by each method over Taxi-2 ($\varepsilon = 1$): (a) DPT [19], (b) AdaTrace [20], (c) DP-MODR [21], (d) LDPTrace [22], and (e) DPTraj-PM.

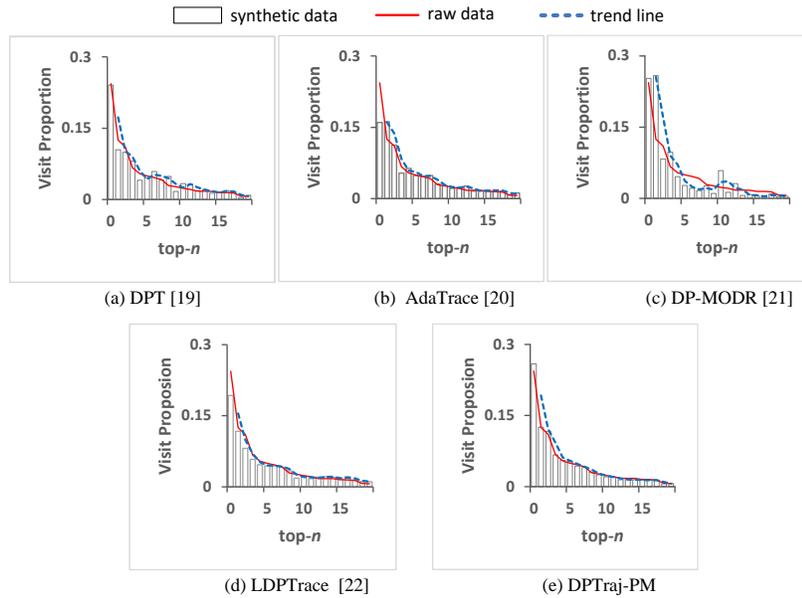

Figure 11: Probability distribution of top-$n$ visited regions for real and synthetic trajectories generated by each method over Geolife ($\varepsilon = 1$): (a) DPT [19], (b) AdaTrace [20], (c) DP-MODR [21], (d) LDPTrace [22], and (e) DPTraj-PM.

DP-MODR [21] builds a pruned noisy cost-sensitive path tree for each cell in order to simulate the most frequent patterns. It may traverse from one noisy cost-sensitive path tree to another without fully consideration of the prior trajectory segment during synthetic trace generation, and thus may lose some data accuracy. Furthermore, only using the most frequent patterns and ignoring some less frequent patterns may also hinder it from preserving more aggregate properties.

DPTraj-PM provides much better data utility than LDPTrace [22]. Due to the use of 1-order Markov mobility model which shares the whole privacy budget with other 2 core components, LDPTrace may not able to retain sufficient movement patterns of the original trajectories, and thus might fail to preserve very desirable data utility.

**5.3 Parameter Analysis**

There are two important parameters in DPTraj-PM: the order $m$ of the Markov process, and the grid size $u_h \times u_w$. Here, we analyze how they affect the data utility. Furthermore, we report the evaluation on the privacy budget allocation for the noisy prefix tree.

*5.3.1 Impact of m on Data Utility.*

Theoretically, if we fix all the parameters of the proposed method except the order $m$ of the Markov process, increasing $m$ values from 1 might first increase the accuracy of the noisy $m$-order Markov Model, and then decrease the accuracy as the number of $(m + 1)$-grams becomes fewer and fewer. The accuracy of the noisy prefix tree might decrease with the increase of $m$ values. Maybe there is a balance between the accuracy of the noisy prefix tree and the noisy $m$-order Markov Model. We plot the results over Geolife with varying $m$ in Figure 12. We find the data utility is better preserved in general when $m = 2, 3$ than $m = 1, 4$, which verifies our analysis above.

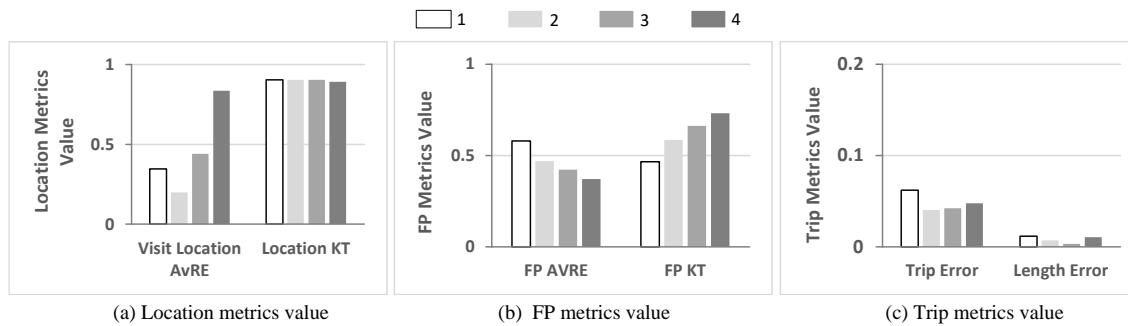

(a) Location metrics value      (b) FP metrics value      (c) Trip metrics value

Figure 12: Data utility preservation performance of DPTraj-PM over Geolife with varying $m$ ($\varepsilon = 1$).

*5.3.2 Impact of $u_h \times u_w$ on Data Utility.*

In the space discretization step, we impose a $u_h \times u_w$ uniform grid over the spatial domain, and map the location points to the anchor points of neighboring cells based on interpolation. This inevitably leads to information loss. The smaller the cells are, the higher accuracy the calibrated trajectories provide. But with the reduction of each cell's size, the number of trajectories that traverse a cell will decrease. For a dataset that is not big enough and/or not evenly distributed (e.g., Taxi and Geolife), this could cause the model to become more sensitive to noise, and hence may lead to the data utility loss of synthetic trajectories. Therefore, to obtain good overall data utility, the grid size should be neither too coarse nor too fine. We plot the results over Taxi-2 with varying grid size in Figure 13. We observe the data utility is better preserved in general when $u_h \times u_w = 20 \times 20$ than $u_h \times u_w = 12 \times 12, 14 \times 14, 16 \times 16, 18 \times 18, 22 \times 22, 24 \times 24$. This verifies our analysis.

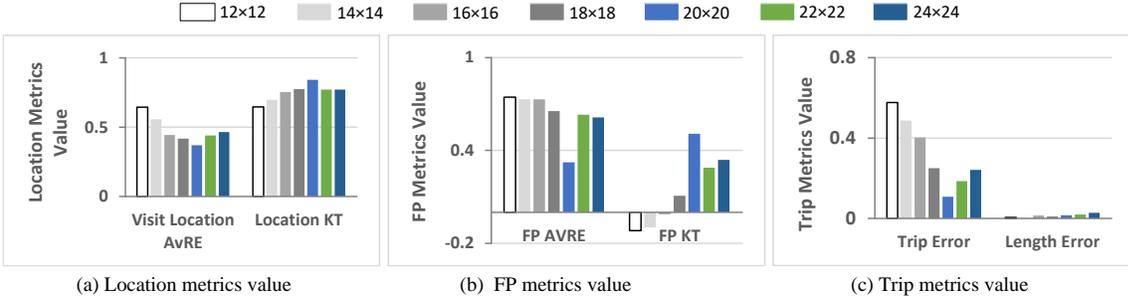

Figure 13: Data utility preservation performance of DPTraj-PM over Taxi-2 with varying grid size ($\varepsilon = 1$, $m = 3$).

#### 5.3.3 Evaluation on Privacy Budget Allocation for Noisy Prefix Tree.

To demonstrate the effectiveness of the decremental privacy budget allocation approach that we exploit for the noisy prefix tree construction, we replace this allocation approach with the uniform allocation [18] and the geometrical allocation [19] in DPTraj-PM, which are respectively referred to as proposed-uniform and proposed-geo, and evaluate their utility on Geolife. The results in Figure 14 indicate that DPTraj-PM that leverages the decremental privacy budget allocation approach performs better than the ones that adopt the uniform allocation or the geometrical allocation.

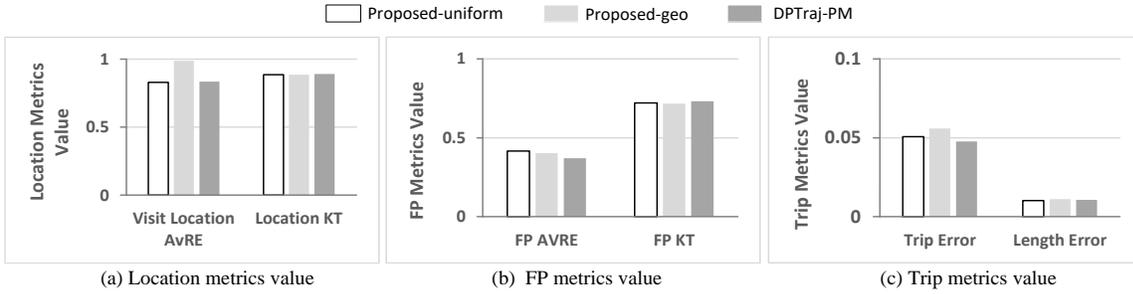

Figure 14: Data utility preservation performance of DPTraj-PM with different privacy budget allocation techniques for the noisy prefix tree construction over Geolife ($\varepsilon = 1$, $m = 4$).

## 6 CONCLUSIONS

In this paper, we present DPTraj-PM, a novel differentially private trajectory synthesizer based on a prefix tree and an $m$-order Markov process. DPTraj-PM uses the prefix tree to simulate the initial trajectory segments, and employs the $m$-order Markov process to pick the next location point. To guarantee privacy protection and preserve data utility, a noise addition approach is carefully designed under differential privacy for it. The experiments on real-life datasets show that DPTraj-PM substantially outperforms state-of-the-art techniques [19-22] in general in terms of data utility and accuracy.

## REFERENCES

[1] Àlex Miranda-Pascual, Patricia Guerra-Balboa, Javier Parra-Arnau, Jordi Forné, and Thorsten Strufe. 2023. SoK: Differentially private publication of trajectory data. In Proceedings on Privacy Enhancing Technologies, 496-516.

[2] Marco Fiore, Panagiota Katsikouliy, Elli Zavouy, Mathieu Cunchey, Franc¸oise Fessantz, Dominique L. Helloz, Ulrich M. Aivodjix, Baptiste Olivierz, Tony Quertierz, and Razvan Stanica. 2020. Privacy in trajectory micro-data: a survey. Transactions on Data Privacy, 13, 91-149.

[3] Yves-Alexandre de Montjoye, Ce´sar A. Hidalgo, Michel Verleysen, and Vincent D. Blondel. 2013. Unique in the crowd: The privacy bounds of

human mobility. Scientific Reports 3, 1, 1-5.

[4] Florimond Houssiau, Luc Rocher, and Yves-Alexandre de Montjoye. 2022. On the difficulty of achieving differential privacy in practice: user-level guarantees in aggregate location data. Nature communications 13, 1, 1-3.

[5] Fengmei Jin, Wen Hua, Matteo Francia, Pingfu Chao, Maria E Orlowska, and Xiaofang Zhou. 2022. A survey and experimental study on privacy-preserving trajectory data publishing. IEEE Transactions on Knowledge and Data Engineering 35, 6, 5577-5596.

[6] Seungjae Shin, Hongseok Jeon, Chunglae Cho, Seunghyun Yoon, and Taeyeon Kim. 2020. User mobility synthesis based on generative adversarial networks: A survey. In 2020 22nd International Conference on Advanced Communication Technology (ICACT). IEEE, 94-103.

[7] Xiangjie Kong, Qiao Chen, Mingliang Hou, Hui Wang, and Feng Xia. 2023. Mobility trajectory generation: a survey. Artificial Intelligence Review 56, S3057–S3098.

[8] Osman Abul, Francesco Bonchi, and Mirco Nanni. 2008. Never walk alone: uncertainty for anonymity in moving objects databases. In IEEE 24th International Conference on Data Engineering. IEEE, 376-385.

[9] Naghizade, Elham, Lars Kulik, Egemen Tanin, and James Bailey. 2020. Privacy-and context-aware release of trajectory data. ACM Transactions on Spatial Algorithms and Systems (TSAS) 6, 1, 1-25.

[10] Vincent Primault, Sonia Ben Mokhtar, Cédric Lauradoux, and Lionel Brunie. 2015. Time distortion anonymization for the publication of mobility data with high utility. In 2015 IEEE Trustcom/BigDataSE/ISPA. IEEE, 539-546.

[11] Jian Kang, Doug Steiert, Dan Lin, and Yanjie Fu. 2019. MoveWithMe: Location privacy preservation for smartphone users. IEEE Transactions on Information Forensics and Security 15, 711-724.

[12] Yuanshao Zhu, Yongchao Ye, Shiyao Zhang, Xiangyu Zhao, and James J.Q. Yu. 2024. Difftraj: Generating GPS trajectory with diffusion probabilistic model. In Advances in Neural Information Processing Systems, 36.

[13] A. Ercument Cicek, Mehmet Ercan Nergiz, and Yucel Saygin. 2014. Ensuring location diversity in privacy-preserving spatio-temporal data publishing. The VLDB Journal—The International Journal on Very Large Data Bases 23, 4, 609-625.

[14] Nana Wang and Mohan Kankanhalli. 2020. Protecting sensitive place visits in privacy-preserving trajectory publishing. Computers & Security 97, 101949.

[15] Carlini Nicolas, Jamie Hayes, Milad Nasr, Matthew Jagielski, Vikash Sehwag, Florian Tramer, Borja Balle, Daphne Ippolito, and Eric Wallace. 2023. Extracting training data from diffusion models. In 32nd USENIX Security Symposium (USENIX Security 23). USENIX Association, 5253-5270. 2023.

[16] Cynthia Dwork, Frank McSherry, Kobbi Nissim, and Adam Smith. 2006. Calibrating noise to sensitivity in private data analysis. In Theory of Cryptography Conference. Springer, 265-284.

[17] Rui Chen, Gergely Acs, and Claude Castelluccia. 2012. Differentially private sequential data publication via variable-length n-grams. In Proceedings of the 2012 ACM conference on Computer and communications security. ACM, 638-649.

[18] Rui Chen, Benjamin C. M. Fung, Bipin C. Desai, and Nériah M. Sossou. 2012. Differentially private transit data publication: a case study on the montreal transportation system. In Proceedings of the 18th ACM SIGKDD International Conference on Knowledge Discovery and Data Mining. ACM, 213-221.

[19] Xi He, Graham Cormode, Ashwin Machanavajjhala, Cecilia M. Procopiuc, and Divesh Srivastava. 2015. DPT: differentially private trajectory synthesis using hierarchical reference systems. In Proceedings of the VLDB Endowment. 1154-1165.

[20] Mehmet Emre Gursoy, Ling Liu, Stacey Truex, Lei Yu, and Wenqi Wei. 2018. Utility-aware synthesis of differentially private and attack-resilient location traces. In Proceedings of the 2018 ACM SIGSAC Conference on Computer and Communications Security. ACM. 196-211.

[21] Fatemeh Deldar and Mahdi Abadi. 2021. Enhancing spatial and temporal utilities in differentially private moving objects database release. International Journal of Information Security 20, 4, 511-533.

[22] Yuntao Du, Yujia Hu, Zhikun Zhang, Ziquan Fang, Lu Chen, Baihua Zheng, and Yunjun Gao. 2023. LDPTrace: Locally differentially private trajectory synthesis. PVLDB 16, 8, 1897 - 1909.

[23] Mehmet Emre Gursoy, Ling Liu, Stacey Truex, and Lei Yu. 2018. Differentially private and utility preserving publication of trajectory data. IEEE Transactions on Mobile Computing 18, 10, 2315-2329.

[24] https://www.kaggle.com/competitions/pkdd-15-predict-taxi-service-trajectory-i/data.

[25] Jun Zhang, Xiaokui Xiao, and Xing Xie. 2016. Privtree: A differentially private algorithm for hierarchical decompositions. In Proceedings of the 2016 International Conference on Management of Data. ACM, 155-170.

[26] Shuo Wang and Richard O. Sinnott. 2017. Protecting personal trajectories of social media users through differential privacy. Computers & Security 67, 142-163.

[27] Khalil Al-Hussaeni, Benjamin C.M. Fung, Farkhund Iqbal, Gaby G. Dagher, and Eun G. Park. 2018. SafePath: Differentially-private publishing of passenger trajectories in transportation systems. Computer Networks 143, 126-139.

[28] Yang Li, Dasen Yang, and Xianbiao Hu. 2020. A differential privacy-based privacy-preserving data publishing algorithm for transit smart card data. Transportation Research Part C: Emerging Technologies 115, 102634.

[29] Jingyu Hua, Yue Gao, and Sheng Zhong. 2015. Differentially private publication of general time-serial trajectory data. In 2015 IEEE Conference on Computer Communications (INFOCOM). IEEE, 549-557.

[30] Xiaodong Zhao, Dechang Pi, and Junfu Chen. 2020. Novel trajectory privacy-preserving method based on clustering using differential privacy. Expert Systems with Applications 149, 113241.

[31] Qi Liu , Juan Yu, Jianmin Han, and Xin Yao. 2021. Differentially private and utility-aware publication of trajectory data. Expert Systems with

Applications 180, 115120.

[32] Quan Geng, Peter Kairouz, Sewoong Oh, and Pramod Viswanath. 2015. The staircase mechanism in differential privacy. IEEE Journal of Selected Topics in Signal Processing 9, 7, 1176–1184.

[33] Darakhshan J. Mir, Sibren Isaacman, Ramon Cáceres, Margaret Martonosi, and Rebecca N. Wright. 2013. Dp-where: Differentially private modeling of human mobility. In 2013 IEEE international conference on big data. IEEE, 580-588.

[34] Harichandan Roy, Murat Kantarcioglu, and Latanya Sweeney. 2016. Practical differentially private modeling of human movement data. In IFIP Annual Conference on Data and Applications Security and Privacy. Springer, Cham, 170-178.

[35] Fengmei Jin, Wen Hua, Boyu Ruan, and Xiaofang Zhou. 2023. Frequency-based randomization for guaranteeing differential privacy in spatial trajectories. In 2022 IEEE 38th International Conference on Data Engineering. IEEE, 1727-1739.

[36] Jing Zhang, Qihan Huang, Yirui Huang, Qian Ding, and Pei-Wei Tsai. 2023. DP-TrajGAN: A privacy-aware trajectory generation model with differential privacy. Future Generation Computer Systems 142, 25-40.

[37] Stella Ho, Youyang Qu, Bruce Gu, Longxiang Gao, Jianxin Li, and Yong Xiang. 2021. DP-GAN: Differentially private consecutive data publishing using generative adversarial nets. Journal of Network and Computer Applications 185, 103066.

[38] Ian Goodfellow, Jean Pouget-Abadie, Mehdi Mirza, Bing Xu, David Warde-Farley, and Sherjil Ozair. 2014. Generative adversarial nets. Advances in Neural Information Processing Systems 27.

[39] Martin Mächler and Peter Bühlmann. 2004. Variable length Markov chains: methodology, computing, and software. Journal of Computational and Graphical Statistics 13, 2, 435-455.

[40] Frank McSherry. 2009. Privacy integrated queries: an extensible platform for privacy-preserving data analysis. In Proceedings of the 2009 ACM SIGMOD International Conference on Management of Data. 19-30.

[41] Cynthia Dwork and Aaron Roth. 2013. The algorithmic foundations of differential privacy. Theoretical Computer Science 9, 3-4, 211-407.

[42] Yu Zheng, Xing Xie and Wei-Ying Ma. 2010. Geolife: A collaborative social networking service among user, location and trajectory. Bulletin of the Technical Committee on Data Engineering 33, 2, 32–39.